\documentclass[aps,twocolumn,pra]{revtex4-1}
\usepackage{amsmath,graphicx,bbm,mathrsfs,amssymb,pst-all,bm,color}

\begin{document}

\title{Quantum algorithm for total least squares data fitting}

\author{Hefeng Wang$^{1}$}
\email{wanghf@mail.xjtu.edu.cn}
\author{Hua Xiang$^{2}$}
\email{hxiang@whu.edu.cn}
\affiliation{$^{1}$Department of Applied Physics, School of Science, Xi'an
Jiaotong University and Shaanxi Province Key Laboratory of Quantum
Information and Quantum Optoelectronic Devices, Xi'an, 710049, China}
\affiliation{$^{2}$School of Mathematics and Statistics, Wuhan University, Wuhan,
430072, China}

\begin{abstract}
The total least squares~(TLS) method is widely used in data-fitting. Compared with the least squares fitting method, the TLS fitting takes into account not only observation errors, but also errors from the measurement matrix of the variables. In this work, the TLS problem is transformed to finding the ground state of a Hamiltonian matrix. We propose quantum algorithms for solving this problem based on quantum simulation of resonant transitions. Our algorithms can achieve at least polynomial speedup over the known classical algorithms.
\end{abstract}

\maketitle

\section{Introduction}

One basic problem in applied mathematics is to create a theoretical model
and make a reliable prediction according to the observed data~\cite%
{VanHuffel}. This problem appears in a broad class of fields such as signal
processing, automatic control, physics, astronomy, biology, statistics,
economics, etc~\cite%
{VanHuffel,VanHuffelLemmerling2002book,Huang_IEEE01,VanHuffel1997}. The
linear model with some parameters is most widely used, and the key is to
determine these unknown parameters from the measurement data of certain
variables. Let $\mathbf{x=}\left( x_{1},\ldots ,x_{N}\right) ^\dag$ be
the parameter vector that characterizes the model. Suppose that the observed
data can be expressed by a linear combination as $a_{i1}x_{1}+\ldots
+a_{iN}x_{N}=b_{i}$, where $a_{i1},\ldots ,a_{iN}$ and $b_{i}$ stand for the
observed data of variables. Such a data-fitting problem usually gives rise
to an overdetermined linear system of equations~with $N$ unknowns $\mathbf{x}
$: $A\mathbf{x}\approx \mathbf{b}$, where $A\in \mathbb{R}^{M\times N}$, $%
\mathbf{x}\in \mathbb{R}^{N}$, $\mathbf{b}\in \mathbb{R}^{M}$ and $M>N$. Our
task is to find an optimal estimate of the parameter vector $\mathbf{x}$
under some constraints.

The well-known least squares~(LS) method is one of the most widely used
methods in data-fitting. It finds the vector $\mathbf{x}$ that minimizes $%
\left\Vert A\mathbf{x}-\mathbf{b}\right\Vert _{2}$, where $\Vert \cdot \Vert
_{2}$ denotes the matrix $2$-norm. The LS solution can be expressed by
Moore-Penrose inverse, and when $A$ is of full column rank, it reads $%
\mathbf{x}_{\text{LS}}=\left( A^\dag A\right) ^{-1}A^\dag \mathbf{b}$.
There exist two types of classical methods for the LS solution~\cite{Drineas06}. The direct method based on QR factorization is the standard one, while the iterative method utilizes the CG-like solvers. The sampling method is recently popular for large scale problems, which reduces the problem size under appropriate assumptions on the sampling probabilities, and achieves an
approximate LS solution efficiently and accurately.

In the LS method, the matrix $A$ is fixed and assumed to be free from error,
while the vector $\mathbf{b}$ is contaminated by errors. However, this
assumption is usually unrealistic in some cases, since the matrix $A$ may
not be accurate as well due to sampling errors, human errors, modeling
errors and instrument errors, etc~\cite[Page 5]{VanHuffel}. The total least
squares~(TLS) method gives a better estimate than the LS method~\cite[Page 5]%
{VanHuffel} when there exist errors in both the vector $\mathbf{b}$ and the
matrix $A$, and especially when these errors are independent random
variables with zero mean and equal variance, i.e., independent and
identically distributed random variables.

Fitting a large amount of data is in fact a difficult task for a classical
computer. For example, the computational complexity of the TLS method
depends on the singular value decomposition~(SVD), which costs about $%
26N^{3} $ by using the R-bidiagonalization~\cite{GolubVanLoan_book13}, where
$N$ is the data size. Even the partial SVD via the $s$-step Lanczos
procedure or randomized SVD needs about $O\left( N^{2}\right) $ flops~\cite%
{VanHuffel,XXW_NLAA18}. Quantum computers can outperform classical computers
in solving a number of problems~\cite{shor, grover, childs, nori}. In Ref.~%
\cite{datafitting}, a quantum data-fitting algorithm was proposed for the
least squares method.

In this work, we propose a quantum algorithm for the TLS method of
data-fitting based on the quantum simulation of resonant transitions. In
Sec.~II, we introduce the TLS fitting method; In Sec.~III, we present a
quantum algorithm for the TLS method, and in Sec.~IV, we simulate the
algorithms through a numerical example, and we conclude in Sec.~V.

\section{Total least squares fitting}

The TLS data fitting problem can be formulated as
\begin{eqnarray}
\left\{ \mathbf{x}_{\text{TLS}},~E_{\text{TLS}},~\mathbf{f}_{\text{TLS}%
}\right\} &:&=\arg \min_{\mathbf{x},E,\mathbf{f}}\left\Vert [E,\ \mathbf{f}%
]~\right\Vert _{F}\text{ }  \notag \\
\text{s.t. }(A+E)\mathbf{x} &=&\mathbf{b}+\mathbf{f},  \label{eqn:DefTLS}
\end{eqnarray}%
where $E$ denotes the errors in the observation matrix $A$, $\mathbf{f}$
denotes the errors in the observation vector $\mathbf{b}$, and $\Vert \cdot
\Vert _{F}$ stands for the Frobenius matrix norm. The TLS method is also
known as errors-in-variables model, measurement error modeling or orthogonal
regression in the statistical literature \cite%
{VanHuffelLemmerling2002book,SchuerMWHuffel_ACA05}. The classical solver for
the TLS problem is based on SVD~\cite{VanHuffel}. For an overdetermined
linear system of equations $A\mathbf{x}\approx \mathbf{b}$, suppose that $A$
and its augmented matrix $C=[A,~\mathbf{b}]$ have SVDs, respectively
\begin{equation*}
S^{\dag }CV=\text{diag}\left( \sigma _{1},\sigma _{2},\ldots ,\sigma
_{N+1}\right) =\Sigma ,
\end{equation*}%
\begin{equation*}
\bar{S}^{\dag }A\bar{V}=\text{diag}\left( \bar{\sigma}_{1},\bar{\sigma}%
_{2},\ldots ,\bar{\sigma}_{N}\right) ,
\end{equation*}%
where $\sigma _{1}\geqslant \bar{\sigma}_{1}\geqslant \sigma _{2}\geqslant
\bar{\sigma}_{2}\ldots \geqslant \bar{\sigma}_{N}\geqslant \sigma _{N+1}$, $%
S $, $\bar{S}$, $V$, and $\bar{V}$ are unitary matrices. In well-conditioned
cases, the genericity condition
\begin{equation}
\bar{\sigma}_{N}>\sigma _{N+1}  \label{eqn:GenericityCond}
\end{equation}%
is satisfied and it ensures the existence and uniqueness of the TLS solution~%
\cite{Golub80}. Let $\mathbf{y=}\left(
\begin{array}{c}
\mathbf{x}_{\text{TLS}} \\
-1%
\end{array}%
\right) $, then the TLS method seeks $C\mathbf{y}\approx \mathbf{0}$ under
the constraint in Eq.~\eqref{eqn:DefTLS}. By solving%
\begin{equation}
\min \mathbf{y}^\dag C^{\dag }C\mathbf{y,}
\end{equation}%
we have $\mathbf{y=}\frac{-\mathbf{v}_{N+1}}{\mathbf{v}_{N+1,N+1}}$, where $%
\mathbf{v}_{N+1}$ is $(N+1)$-th column vector of the matrix $V$ associated
with the corresponding singular value $\sigma _{N+1}$, and $\mathbf{v}%
_{N+1,N+1}$ is the $(N+1)$-th component of $\mathbf{v}_{N+1}$. Then the TLS
solution to the problem is $\mathbf{x}_{\text{TLS}}=\frac{-1}{\mathbf{v}%
_{N+1,N+1}}[\mathbf{v}_{1,N+1},~\ldots ,\mathbf{v}_{N,N+1}]^T$ \cite%
{VanHuffelLemmerling2002book}. That is, it can be obtained by finding the
ground state $\mathbf{v}_{N+1}$ of the matrix $C^{\dag }C$. The TLS solution
can also be expressed as%
\begin{equation}
\mathbf{x}_{\text{TLS}}=\left( A^{\dag }A-\sigma _{N+1}^{2}I\right)
^{-1}A^{\dag }\mathbf{b,}
\end{equation}%
following Ref.~\cite[Theorem 2.7]{VanHuffel}.

\section{Quantum algorithms for TLS fitting}

From the above analysis, we can see that obtaining TLS solution to a
data-fitting problem can be reduced to finding the singular vector $\mathbf{v%
}_{N+1}$, which corresponds to the smallest singular value of the augmented
matrix $C$. The vector $\mathbf{v}_{N+1}$ can be obtained by solving the
eigenproblem of an extended matrix $\left(
\begin{array}{cc}
0 & C \\
C^{\dag } & 0%
\end{array}%
\right) $~\cite{HHL}, whose eigenvalues are $\left\{ \pm \sigma _{j}\right\}
$, with the corresponding eigenstates proportional to states $(\mathbf{s}%
_{j},\pm \mathbf{v}_{j})$, where $\mathbf{s}_{j}$ are column vectors of the
matrix $S$. For simplicity, we use a Hermitian matrix
\begin{equation}
D\mathbf{=}C^{\dag }C,
\end{equation}%
whose eigenvalues and the corresponding eigenvectors are $\sigma _{j}^{2}$
and $\mathbf{v}_{j}$, respectively. The approach we introduced can be
applied for the extended matrix directly. The problem of TLS fitting is
transformed to finding the ground state of the matrix $D$. In this work, we
propose quantum algorithms based on the quantum simulation of resonant
transitions and apply them for the TLS fitting.

We have proposed quantum algorithms for solving eigenproblems of a physical
system~\cite{WHF2016, WHF2017}. When the transition frequency between two
energy levels of the system matches the frequency of a probe qubit coupled
to the system, the probe qubit exhibits a dynamical response. By varying the
frequency of the probe qubit and identifying the locations of resonance
peaks, the energy spectrum of the system can be determined. And the system
can be controllably evolved to the eigenstate with any desired eigenvalue.
Therefore the energy spectrum and the corresponding eigenstates of the
system can be obtained. In this work, we optimize the previous algorithms
and apply them for TLS fitting. The present algorithm requires $\left(
n+1\right) $ qubits, with one probe qubit and $n$ qubits representing the
matrix $D$, and $2^{n}\geqslant N+1$. Details of the algorithms are
described as follows.

\subsection{Algorithm I}

The Hamiltonian of the algorithm is constructed as%
\begin{equation}
H^{\left( 1\right) }=-\frac{1}{2}\omega \sigma _{z}\otimes I+H_{R}^{\left(
1\right) }+c\sigma _{x}\otimes F,
\end{equation}%
where
\begin{equation}
H_{R}^{\left( 1\right) }=\varepsilon _{0}|1\rangle \langle 1|\otimes |\psi
\rangle \langle \psi |+|0\rangle \langle 0|\otimes D\mathbf{,}
\end{equation}%
and $I$ is the identity operator, $\sigma _{x,z}$ are the Pauli matrices,
and $\omega $ is the frequency of the probe qubit. The first term in Eq.~($6$%
) is the Hamiltonian of the probe qubit, the matrix $D$ is encoded in the
second term as shown in Eq.~($7$), and the third term describes the coupling
between the probe qubit and the $n$-qubit quantum register that represents
the system. Here, $\varepsilon _{0}$ is a reference parameter, and $c\ll 1$
is the coupling strength. In the algorithm, we set the probe qubit in its
excited state $|1\rangle $ and the $n$-qubit register in a reference state $%
|\psi \rangle $ that can be easily prepared, then the initial state of the
circuit is an eigenstate of $H_{R}^{(1)}$ with eigenvalue $\varepsilon _{0}$%
. The operator $F$ acts as a transition operator that transforms the
reference state $|\psi \rangle $ to an eigenstate of the matrix $D$ with the
desired eigenvalue. Its form depends on the eigenstate of interest. We give
a detailed form of the operator $F$ for TLS fitting in Sec.~III.C.

In order to obtain the eigenstate $|\mathbf{v}_{N+1}\rangle $ of the matrix $%
D$, we need to obtain its corresponding eigenvalue first. By varying the
frequency of the probe qubit $\omega $ or the reference parameter $%
\varepsilon _{0}$, we can locate the transition frequencies between the
reference state and the eigenstates of $D$ through the quantum simulation of
resonant transitions, thus the ground state eigenvalue of the matrix $D$ can
be obtained. The procedures of the algorithm are briefly summarized as
follows.

First, we make a guess on the range of the ground state eigenvalue of the
matrix $D$ as $\left[ \lambda _{\min }\text{, }\lambda _{\max }\right] $.
Set the reference parameter $\varepsilon _{0}<\lambda _{\min }$, and the
transition frequency between the reference state and the eigenstates of $D$
is $\left[ \omega _{\min }\text{, }\omega _{\max }\right] $ with $\omega
_{\min }=\lambda _{\min }-\varepsilon _{0}$ and $\omega _{\max }=\lambda
_{\max }-\varepsilon _{0}$~(Here we use atomic units). We discretize this
frequency range into $l$ intervals, where each interval has a width of $%
\Delta \omega =\left( \omega _{\max }-\omega _{\min }\right) /l$, and the
frequencies are given by $\omega _{k}=\omega _{\min }+k\Delta \omega
,k=0\ldots ,l-1$. We run the following steps by scanning these frequency
points of the probe qubit: first we prepare the initial state of the circuit
as $|1\rangle |\psi \rangle $, such that $H_{R}^{(1)}|1\rangle |\psi \rangle
=\varepsilon _{0}|1\rangle |\psi \rangle $; then set the transition
frequency of the probe qubit $\omega =\omega _{k}$ and evolve the circuit
with the Hamiltonian $H^{\left( 1\right) }$ for time $t$ by implementing the
time evolution operator $U^{\left( 1\right) }=\exp \left( -iH^{\left(
1\right) }t\right) $; after that, perform a measurement on the probe qubit
in its computational basis. Repeat these steps for a number of times to
obtain the decay probability of the probe qubit until run over all the
frequency points.

After obtaining the ground state eigenvalue $\sigma _{N+1}^{2}$ of the
matrix $D$, we encode this value in the Hamiltonian of the algorithm by
setting $\varepsilon _{0}$ and $\omega $ such that $\sigma
_{N+1}^{2}-\varepsilon _{0}=\omega =1$, which is the condition for the
resonant transition to occur. Then run the algorithm again, once the probe
qubit is observed to decay to its ground state $|0\rangle $, it indicates
that the $n$-qubit register collapse to the ground state $|\mathbf{v}%
_{N+1}\rangle $.

In this algorithm, the operator $F$ acts on the state $|\psi \rangle $ and
can be spanned by the complete set of eigenstates of the matrix $D$ as $%
F|\psi \rangle =\sum\nolimits_{i=1}^{N+1}g_{i}|\mathbf{v}_{i}\rangle $,
where $g_{i}=\langle \mathbf{v}_{i}|F|\psi \rangle $. In basis \{$|1\rangle
|\psi \rangle $, $|0\rangle |\mathbf{v}_{i}\rangle $, $i=1$, $2$, $\ldots $,
$N+1$\}, with the condition $\sigma _{N+1}^{2}-\varepsilon _{0}=\omega =1$,
the resonant transition between states $|1\rangle |\psi \rangle $ and $%
|0\rangle |\mathbf{v}_{N+1}\rangle $ is induced since $%
H_{00}^{(1)}=H_{N+1,N+1}^{(1)}$. The system evolves from the initial state
to the state $|0\rangle |\mathbf{v}_{N+1}\rangle $ reaches its maximal
probability at time $t\sim \pi /\left( 2c|\langle \mathbf{v}_{N+1}|F|\psi
\rangle |\right) $, provided that $\sigma _{N}^{2}-\sigma _{N+1}^{2}\gg c$.
Ignoring the off-resonant transitions, the decay probability of the probe
qubit can be approximated as $\sin ^{2}\left( \frac{Qt}{2}\right) $, where $%
Q=2c|\langle \mathbf{v}_{N+1}|F|\psi \rangle |$. If $|\langle \mathbf{v}%
_{N+1}|F|\psi \rangle |$ is finite, the system can be evolved to the state $|%
\mathbf{v}_{N+1}\rangle $ in finite time.

\subsection{Algorithm II}

We can also obtain the ground state of the matrix $D$ through projection
from an initial guess state and purify it using an iterative procedure based
on a resonance mechanism. The Hamiltonian of the second algorithm is
constructed as
\begin{equation}
H^{\left( 2\right) }=-\frac{1}{2}\omega \sigma _{z}\otimes I+H_{R}^{\left(
2\right) }+c\sigma _{x}\otimes I,
\end{equation}%
where
\begin{equation}
H_{R}^{\left( 2\right) }=\varepsilon _{0}|1\rangle \langle 1|\otimes
I+|0\rangle \langle 0|\otimes D.
\end{equation}%
In this algorithm, we first make a guess on the state $|\mathbf{v}%
_{N+1}\rangle $ as $|\varphi ^{(0)}\rangle $. As in algorithm I, we need to
obtain the ground state eigenvalue of the matrix $D$ first, then find the
ground state $|\mathbf{v}_{N+1}\rangle $.

By preparing the initial state of the circuit as $|1\rangle |\varphi
^{(0)}\rangle $ and the time evolution operator $U^{\left( 2\right) }=\exp
\left( -iH^{\left( 2\right) }\tau \right) $ with $\tau =\pi /(2c)$, the
ground state eigenvalue of the matrix $D$ can be obtained using the same
procedures as in algorithm I. Then we can encode the ground state eigenvalue
$\sigma _{N+1}^{2}$ in the Hamiltonian of the algorithm by setting $%
\varepsilon _{0}$ and $\omega $ such that $\sigma _{N+1}^{2}-\varepsilon
_{0}=\omega =1$. Then we can run the following procedures to obtain the
state $|\mathbf{v}_{N+1}\rangle $: First, we prepare the circuit in state $%
|1\rangle |\varphi ^{(0)}\rangle $; then implement the time evolution
operator $U^{\left( 2\right) }$; next perform a measurement on the probe
qubit in its computational basis. A measurement is defined as
\textquotedblleft successful measurement\textquotedblright\ only if the
measurement result on the probe qubit is in its ground state $|0\rangle $.
If a successful measurement is performed, then set the probe qubit in state $%
|1\rangle $ and implement the time evolution operator $U^{\left( 2\right) }$
again, and perform a measurement on the probe qubit. Repeat these steps
until $j$ continuous successful measurements are achieved. The state $%
|\varphi ^{(j)}\rangle $ obtained on the last $n$ qubits of the circuit is
close to the ground state $|\mathbf{v}_{N+1}\rangle $ of $D$.

The state $|\varphi ^{(0)}\rangle $ can be spanned by the eigenstates of the
matrix $D$ as $|\varphi ^{(0)}\rangle =\sum\nolimits_{i=1}^{N+1}d_{i}|%
\mathbf{v}_{i}\rangle $, where $d_{i}=\langle \varphi ^{(0)}|\mathbf{v}%
_{i}\rangle $ and $\sum\nolimits_{i=1}^{N+1}|d_{i}|^{2}=1$. In basis \{$%
|1\rangle |\mathbf{v}_{i}\rangle $, $|0\rangle |\mathbf{v}_{i}\rangle $, $%
i=1 $, $\ldots $, $N+1$\}, with the condition $\sigma _{N+1}^{2}-\varepsilon
_{0}=\omega =1$, resonant transition between states $|1\rangle |\varphi
^{(0)}\rangle $ and $|0\rangle |\mathbf{v}_{N+1}\rangle $ is induced. As
analyzed in Ref.~\cite{WHF2017}, the success probability of achieving $j$
continuous successful measurements on the probe qubit is proportional to $%
|\langle \varphi ^{(0)}|\mathbf{v}_{N+1}\rangle |^{2}$. As long as $|\langle
\varphi ^{(0)}|\mathbf{v}_{N+1}\rangle |^{2}$ is finite, and the gap $\sigma
_{N}^{2}-\sigma _{N+1}^{2}$ between the ground state and the first excited
state of $D$ is not exponentially small and $\gg c$, the system converges
quickly to the state $|\mathbf{v}_{N+1}\rangle $. The system is evolved to
its ground state in polynomial time with polynomial large success
probability.

\subsection{Application of the algorithms for TLS fitting}

We have to provide a good initial guess on the ground state of the matrix $D$
in order to run the algorithms efficiently. In TLS fitting, the errors
introduced in the measurement matrix $A$ are independent and identically
distributed random variables. It is reasonable to assume that the effects of
these noises are small and can be treated perturbatively. Thus the LS
solution of the fitting problem should be a good initial guess to that of
the TLS solution. Mathematically, comparing the TLS solution shown in Eq.~($%
4 $) with the LS solution $\mathbf{x}_{\text{LS}}=\left( A^{\dag }A\right)
^{-1}A^{\dag }\mathbf{b}$ of a fitting problem, we can see that in the TLS
solution, the contribution from the smallest singular value of the augmented
matrix $C$ is taken into account.
When the singular value $\sigma _{N+1}$ is sufficiently small, it is
reasonable to assume that the LS solution provides a good approximation to
the TLS solution of the fitting problem. This can also be derived as
follows. From Eq.~($4$), assuming that $A$ is of full column rank, we have%
\begin{eqnarray}
\mathbf{x}_{\text{TLS}} &=&\left( A^{\dag }A-\sigma _{N+1}^{2}I\right)
^{-1}A^{\dag }\mathbf{b}  \notag \\
&=&\left[ I-\sigma _{N+1}^{2}\left( A^{\dag }A\right) ^{-1}\right]
^{-1}\left( A^{\dag }A\right) ^{-1}A^{\dag }\mathbf{b}  \notag \\
&=&\left[ I-\sigma _{N+1}^{2}\left( A^{\dag }A\right) ^{-1}\right] ^{-1}%
\mathbf{x}_{\text{LS}}.
\end{eqnarray}%
Thus%
\begin{equation}
\mathbf{x}_{\text{TLS}}-\mathbf{x}_{\text{LS}}=\sigma _{N+1}^{2}\left(
A^{\dag }A\right) ^{-1}\mathbf{x}_{\text{TLS}}.
\end{equation}%
Then we get
\begin{equation}
\frac{\left\Vert \mathbf{x}_{\text{TLS}}-\mathbf{x}_{\text{LS}}\right\Vert
_{2}}{\left\Vert \mathbf{x}_{\text{TLS}}\right\Vert _{2}}\leqslant \sigma
_{N+1}^{2}\left\Vert \left( A^{\dag }A\right) ^{-1}\right\Vert _{2}=\left(
\frac{\sigma _{N+1}}{\bar{\sigma}_{N}}\right) ^{2}.
\end{equation}%
We can see that $\mathbf{x}_{\text{LS}}$ is a good approximation to $\mathbf{%
x}_{\text{TLS}}$ when $\sigma _{N+1}\ll \bar{\sigma}_{N}$.

To resolve the TLS solution of a fitting problem, in algorithm I, we set the
initial state of the system as $|\mathbf{b}\rangle $, where $|\mathbf{b}%
\rangle $ is the normalized vector of $\mathbf{b}$. The operator $%
H_{R}^{\left( 1\right) }$ in Eq.~($7$) is set as $H_{R}^{\left( 1\right)
}=\varepsilon _{0}|1\rangle \langle 1|\otimes |\mathbf{b}\rangle \langle
\mathbf{b}|+|0\rangle \langle 0|\otimes D$, and the operator $F$ in Eq.~($6$%
) is set as $F=\left( A^{\dag }A\right) ^{-1}A^{\dag }$. The transition
element between the initial state $|\mathbf{b}\rangle $ and the ground state
of the matrix $D$ is $\langle \mathbf{v}_{N+1}|F|\mathbf{b}\rangle $. The
state $F|\mathbf{b}\rangle $ is proportional to the normalized LS solution $|%
\mathbf{x}_{\text{LS}}\rangle $ of the problem up to a normalized factor.
The evolution time of the algorithm scales as $\pi /\left( 2c|\langle
\mathbf{v}_{N+1}|F|\mathbf{b}\rangle |\right) $. The Moore-Penrose
pseudoinverse operator $\left( A^{\dag }A\right) ^{-1}A^{\dag }$ can be
implemented using the approach introduced in Ref.~\cite{datafitting}. Here
we assume that the matrix $A$ is Hermitian, if $A$ is not Hermitian, we can
construct an extended matrix of $A$ and implement the Moore-Penrose
pseudoinverse of the extended matrix.

In algorithm II, we can first run the quantum algorithm for LS fitting~\cite%
{datafitting} to obtain the LS solution $\left( A^{\dag }A\right)
^{-1}A^{\dag }|\mathbf{b}\rangle $ of the fitting problem, then use it as
the initial state of the algorithm to resolve the TLS solution of the
problem. The TLS solution can be projected out with a probability
proportional to $|\langle \mathbf{v}_{N+1}|\mathbf{x}_{\text{LS}}\rangle
|^{2}$. From the above analysis, we can see that in algorithm I, the LS
solution is encoded implicitly in the algorithm, while in algorithm II, the
LS solution is used explicitly as initial input state of the algorithm.

The runtime of our algorithms consists of two parts: the number of
experiments that needs to be performed to obtain eigenvalue spectrum in a
given scanned eigenvalue range, and the time needed to run the circuit in
each experiment. In the first part, the number of experiments scales as $O(1/\epsilon ^{2})$ where $\epsilon $ denotes accuracy of the ground state
eigenvalue of the matrix $D$. The second part depends on the evolution time
of the algorithm and the computational cost for simulating the algorithm
Hamiltonians as shown in Eqs.~($6$) and ($8$), i.e. implementing the time
evolution operator $e^{-iH^{(1)}t}$ and $e^{-iH^{(2)}t}$. The evolution time
of the algorithm scales as $\pi /\left( 2c|\langle \mathbf{v}_{N+1}|F|\psi
\rangle |\right) $ in algorithm I, and it will be finite as long as $|\langle \mathbf{v}_{N+1}|F|\psi \rangle |$ is finite. In algorithm II, the
TLS solution is projected out with a probability proportional to $|\langle
\mathbf{v}_{N+1}|\mathbf{x}_{\text{LS}}\rangle |^{2}$. As long as the
overlap $|\langle \mathbf{v}_{N+1}|\mathbf{x}_{\text{LS}}\rangle |^{2}$ is
polynomially large, the cost of the algorithm will be polynomial. In
general, the LS solution can serve as a good initial guess for the
corresponding TLS solution of the fitting problem, therefore the evolution
time of the algorithms is finite. The time evolution operators of the
algorithms can be implemented through the Trotter formula~\cite{nc}. There
are also other algorithms for Hamiltonian simulation such as Taylor series
approach~\cite{Berry-sparsematrix} and quantum signal processing~\cite{chuang}. For sparse matrices, they can be simulated in an amount of time
that grows nearly linearly with its sparsity~\cite{AT, Berry1}. For dense
matrices, it was shown in~\cite{BC} that given black-box access to the
matrix elements, Hamiltonian simulation with an error $\delta _{h}$ can be
performed in time $O\left( N^{2/3}poly\log \left( N\right) /\delta
_{h}^{1/3}\right) $. In~Ref.~\cite{WZP}, instead of assuming black-box
access to the matrix elements, they propose to use a memory model where the
entries of the matrices are stored in a data structure in a quantum random
access memory~(qRAM)~\cite{GLM1}. The algorithm achieves
sparsity-independent runtime scaling of $O\left( \kappa ^{2}\sqrt{N}\text{poly}\log \left( N\right) /\epsilon \right) $, where $\kappa $ is the
condition number and $\epsilon $ is the precision to which the solution is
approximated. The Hamiltonians of our algorithm can be calculated directly
and simulated using these algorithms. Based on the above analysis, compare
with the classical TLS fitting algorithms, our algorithms can achieve at
least polynomial speedup in general.

In our algorithm, we obtain the quantum state $|\mathbf{x}_{\text{TLS}}\rangle $, which contains the solution to the TLS problem but different
from the classical one, since one has to measure the state to learn the
solution. For some problems in machine learning, such as big data
classification, the quantum state $|\mathbf{x}_{\text{TLS}}\rangle $ can be
used directly. In many cases, the final output involves dimensionality
reduction or compression, and we extract useful global information, rather
than directly accessing each component of the state~\cite{RL}. For example,
in quantum classifier, the TLS solution can be encoded in a quantum state $|\mathbf{x}_{\text{TLS}}\rangle $. Given a query state, we classify it as $+1$
or $-1$ (output) by performing a SWAP test with the quantum state $|\mathbf{x}_{\text{TLS}}\rangle $ and measuring the success probability~\cite{RL1}.

We can learn the TLS solution of the problem from the quantum state $|\mathbf{x}_{\text{TLS}}\rangle $ using the method in Ref.~\cite{datafitting}
when it is necessary. We can find the concise representation of fit
functions that approximates the vector $\mathbf{b}$ within a certain error
by using quantum state tomography and statistical sampling. In fact, one can
choose the most important $M^{\prime }$ fit functions, where $M^{\prime }$
scales as $poly(\log N)$, and prepare the state $|\mathbf{x}_{\text{TLS}}\rangle $ using compressed sensing. The fitting parameters for the $M^{\prime }$ fit functions in the state $|\mathbf{x}_{\text{TLS}}\rangle $
can be evaluated using SWAP test. We can also estimate the fitting quality
of the parameters $|\mathbf{x}_{\text{TLS}}\rangle $ efficiently by
estimating the quantity $\left\vert \left\langle \mathbf{b}\left\vert
\mathbf{I}\left( A\right) \right\vert \mathbf{x}_{\text{TLS}}\right\rangle
\right\vert ^{2}$, where $\mathbf{I}\left( A\right) =\left(
\begin{array}{cc}
0 & A \\
A^{\dag } & 0\end{array}\right) $, as in Ref.~\cite{datafitting}. The SWAP test is used to determine
the accuracy by performing a SWAP operation on the two quantum states $|\mathbf{b}\rangle $ and $|\mathbf{I}(A)\mathbf{x}_{\text{TLS}}\rangle $
controlled by an auxiliary qubit. There exist the overheads of extracting
such useful classical information. If compressed sensing technique is used
to reconstruct the state within error $O(\epsilon )$, we need $O({M^{\prime }}^{2}\log {M^{\prime }}^{2}/\epsilon ^{2})$ measurements. The SWAP test for
quality estimate needs $O(T_{\text{H}}/\delta ^{2})$, where $T_{\text{H}}$
is the cost for Hamiltonian simulation and $\delta $ is the accuracy of
estimation~\cite{datafitting}.

\section{Numerical simulation of the algorithms}

In the following, we simulate the algorithms through a numerical example in
linear prediction~(LP). We consider the time series expressed by $p$
sinusoids in the form $s(t)=\sum_{j=1}^{p}c_{j}e^{\lambda _{j}t}$, where the
$\lambda _{j}$'s and $c_{j}$'s are to be determined. The $\lambda _{j}$'s
are fundamental constants representing the natural decay of electromagnetic,
acoustic and mechanical systems, while the $c_{j}$'s depend upon the
excitation, sensor location, and time origin, etc. Let $z_{j}=\exp \left(
\lambda _{j}T\right) $,$~j=1,\ldots ,p$, where $T$ is a constant of the
sample rate. Then we have the approximation of complex valued data $%
\{s_{k}\} $ in the form $s_{k}=\sum_{j=1}^{p}c_{j}z_{j}^{k}$. Such linear
sum of damped complex exponentials with uniformly samples is widely used in
Prony analysis, and it is essentially a nonlinear fitting problem for the
amplitudes and frequencies.

We consider an LP model $\widehat{s}_{k}=\sum_{i=0}^{N-1}\alpha
_{i}s_{k-N+i} $, where $s_{k-j}$ ($j=1,\cdots ,N;k\geqslant j$) are previous
observed values, and $\widehat{s}_{k}$ the predicted signal value. That is,
a linear function of previous samples are used to estimate the future
values. Let $A_{N}=\left[ a_{1},\ldots ,a_{N}\right] ,~b_{N}=-a_{N+1}$,
where $a_{j}=\left[ s_{j-1},\ldots ,s_{j+M-2}\right] ^{\mathrm{T}}$. The
linear prediction (LP) equation is formed with the unknown predictor
coefficient vector $\mathbf{x}=\left( \alpha _{0},\alpha _{1},\ldots ,\alpha
_{N-1}\right) ^{\mathrm{T}}$, which is achieved by solving the linear system
\begin{equation}
A_{N}\mathbf{x}=b_{N}.  \label{eqn:PronyLinSys}
\end{equation}%
We can check that $A_{N}$ is a Hankel matrix and $\text{rank}(A_{N})=\min
\{N,p\}$. If $N\geq p$, then the linear system \eqref{eqn:PronyLinSys} is
compatible. For any solution, we construct the characteristic equation
\begin{equation*}
p_{N}(t)=\alpha _{0}+\alpha _{1}t+\cdots +\alpha _{N-1}t^{N-1}+t^{N}.
\end{equation*}%
We know that $p_{N}(t)$ always contains $z_{1},\ldots ,z_{p}$ as its zeros
when $N\geq p$, and hence the frequency can be derived from the roots of the
characteristic equation. Then the amplitudes can be solved from the set of
observed samples that are linear in amplitudes.

Here we only focus on the linear system~\eqref{eqn:PronyLinSys} and its TLS
solution. For the set of LP equations, both the coefficient matrix and the
vector are contaminated by noises.
It is natural to use the TLS solver, which is a promising method in signal
processing. For example, Rahman and Yu~\cite{RahmanYu_IEEE87} presented a
method for frequency and amplitude estimations by using TLS to solve the LP
equations. We choose $c_{j}$ and $\lambda _{j}$ as in Table~\ref{tab:prony},
where $p=12$. The parameters $T=0.2$, $N=11$ and $M=2^{8}$ are used. This
test problem is taken from~\cite{Blaricum_IEEE78,Wei,XXW_NLAA18}, which is
regarded as a benchmark for TLS fitting.

\begin{table}[tbp]
\caption{The parameters in LP equation. }
\label{tab:prony}\centering
\begin{tabular}{ccccc}
\hline
$\lambda_j$ & $c_j$ & \qquad\qquad & $\lambda_j$ & $c_j$ \\
\cline{1-2}\cline{4-5}
$-0.082 \pm 0.926 i$ & 1 &  & $-0.220 \pm 6.800i$ & 1 \\
$-0.147 \pm 2.874 i$ & 1 &  & $-0.247 \pm 8.767i$ & 1 \\
$-0.188 \pm 4.835i$ & 1 &  & $-0.270 \pm 10.733i$ & 1 \\ \hline
\end{tabular}%
\end{table}

We simulate the algorithm I for solving this TLS fitting problem by setting $%
\varepsilon _{0}=-1.0$, $c=0.0005$, and the evolution time $t=30000$. The
initial state of the system is set as $|\mathbf{b}\rangle $, which is the
normalized vector of $\mathbf{b}$, and the operator $F$ is set as $F=\left(
A^{\dag }A\right) ^{-1}A^{\dag }$. By varying the transition frequency $%
\omega $ of the probe qubit and running the algorithm, we obtain the
transition frequency spectrum between the reference state and the ground
state of the matrix $D$ as shown in Fig.~$1$. The ground state eigenvalue of
the matrix $D$ was obtained as $0.0046$. At $\omega =1.0046$, we obtain a
state whose fidelity deviate only in order of $10^{-13}$ from the ground
state of the matrix $D$.
\begin{figure}[tbp]
\includegraphics[width=0.9\columnwidth, clip]{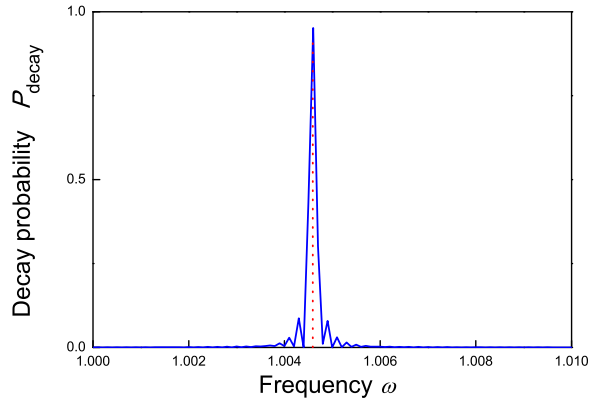}
\caption{(Color online)~Transition frequency spectrum between the reference
state $|\mathbf{b}\rangle $ and the ground state of the matrix $D$ by
simulating algorithm I. The blue solid curve represents the decay
probability of the probe qubit at different frequencies with the coupling
coefficient $c=0.0005$ and the evolution time $t=30000$, the reference
parameter is set as $\protect\varepsilon _{0}=-1.0$. The red dotted vertical
lines represent the known transition frequencies between the reference state
and the eigenstates of the matrix $D$.}
\end{figure}
In simulating algorithm II, we set $c=0.0001$ and $\tau =\pi /(2c) = 15708$.
The initial state of the system is set as $|\mathbf{x}_{\text{LS}}\rangle $,
which is the normalized LS solution $\mathbf{x}_{\text{LS}}$ of the fitting
problem. The transition frequency spectrum of the ground state of the matrix
$D$ with respect to the reference state are obtained as shown in Fig.~$2$.
After running the algorithm again for one iteration by setting $\omega
=1.0046$, the state we obtained has fidelity that deviates from the TLS
solution of the problem only in order of $10^{-12}$. In this example, the
overlap between the LS and TLS solutions of the fitting problem, thus the
success probability of the algorithm at $\omega =1.0046$ is about $0.998$.
The LS solution is very close to the TLS solution of the fitting problem.
This can also be predicted from the eigenvalue spectrum of the matrix $D$,
the ground state eigenvalue of $D$ is $0.0046$, while the eigenvalue of the
first excited state is about $0.908$, which is much larger than the ground
state eigenvalue. In this case, we can see that it is reasonable to use LS
solution as an initial guess for TLS problem. From the numerical simulation,
we can see that by introducing a resonance mechanism, our algorithm can
evolve the initial state quickly to the TLS solution with very high
accuracy.
\begin{figure}[tbp]
\includegraphics[width=0.9\columnwidth, clip]{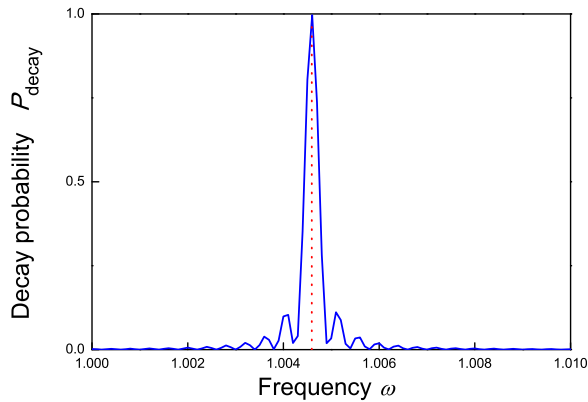}
\caption{(Color online)~Transition frequency spectrum between the reference
state, $|\mathbf{x}_{LS}\rangle $, and the two lowest eigenstates of the
matrix $D$ by simulating algorithm II. The blue solid curve represents the
decay probability of the probe qubit at different frequencies with the
coupling coefficient $c=0.0001$, the evolution time $\protect\tau =15700$,
and the reference parameter $\protect\varepsilon _{0}=-1.0$. The red dotted
vertical lines represent the known transition frequencies between the
reference state and the eigenstates of the matrix $D$.}
\end{figure}

\section{Discussion}

The TLS fitting method takes into account the errors introduced in the
measurement matrix together with those in the observation vector. In the
areas such as signal processing, and geophysics, etc., the TLS method is
more practical than the LS method, since the measurement matrix $A$ and
observation vector $\mathbf{b}$ are both contaminated by noises. In the
generic case, the TLS method yields a unique solution, which is given in
analytic form in terms of the singular value decomposition of the augmented
data matrix $C=[A,\mathbf{b}]$. Precisely speaking, the TLS solution is
expressed by the right singular vector associated with the smallest singular
value of $C$, which corresponds to the eigenvector associated with the
smallest eigenvalue of the Hermitian matrix $D=C^{\dag }C$. The TLS
data-fitting method is then transformed to finding the the ground state of a
Hermitian matrix. We presented two algorithms based on the quantum
simulation of resonant transitions to solve this problem. In our algorithms,
any desired eigenstate of a system can be obtained by inducing proper
resonant transitions between a probe qubit and a transition in the system
that is simulated on a quantum computer. We show that in general, the LS
fitting method can be a good approximation to the TLS solution, and thus can
be used as the initial guess state in the algorithms. This work can be
further generalized to the truncated TLS solution for inverse problems, and
this will be our future work.

Adiabatic quantum evolution~(AQE)~\cite{lidar} algorithms and the
PEA can also be applied for the TLS problem, which is transformed to finding
the ground state of a Hermitian matrix, and both algorithms can solve this
problem. In AQE, starting from an initial Hamiltonian and its ground state,
the system is evolved adiabatically to the target Hamiltonian and its ground
state. The runtime depends on the minimum energy gap between the ground and
the first excited states of the time-dependent adiabatic evolution
Hamiltonian. In our algorithm, the system is evolved to the ground state of
the problem Hamiltonian matrix directly through quantum simulation of
resonant transitions. It requires only information about the spectrum of the
Hamiltonian matrix, without implementing the time-dependent
adiabatic evolution Hamiltonian. Besides, whether the system is evolved to
its ground state is heralded by the non-invasive measurements on the probe
qubit.

The PEA randomly obtains one of the eigenstates of a quantum system from an initial guess state and produces its eigenvalue. The success probability for
obtaining a given eigenstate is proportional to the overlap between the the initial
guess state and the desired eigenstate. In solving the TLS problem by using the PEA,
one can use the quantum state obtained from the LS data-fitting algorithm as
an initial guess state, and apply PEA to project out the TLS solution from
the LS solution. In algorithm I, the combination $F|\psi \rangle $ plays a
somewhat similar role to the initial guess state in the PEA. In our
algorithm, once a transition frequency is identified by running the
algorithm, the algorithm can be repeated with the parameters set to the
resonance condition and the transition to the specific eigenstate occurs
deterministically. Future runs of the algorithm can use this information to
deterministically induce the transition to prepare the corresponding
eigenstate. Also, our algorithm is simpler in that the transitions are
heralded by the state of a single qubit obtained through measurements on the
probe qubit, this makes the observation easier.

\section*{Acknowledgement}

H. Wang is supported by the National Natural Science Foundation of China
(Grant No.~11275145) and the Natural Science Fundamental Research Program of
Shaanxi Province of China under grants 2018JM1015. H. Xiang is supported by
the Natural Science Foundation of China under grants 11571265 and NSFC-RGC
No.~11661161017.

\end{document}